\documentstyle[12pt,aaspp]{article}
\def\ltsima{$\; \buildrel < \over \sim \;$}
\def\lsim{\lower.5ex\hbox{\ltsima}}
\def\gtsima{$\; \buildrel > \over \sim \;$}
\def\gsim{\lower.5ex\hbox{\gtsima}}
\def\bi#1{\bbox{#1}}

\begin{document}
\title{Cosmological Model Predictions for Weak Lensing: 
Linear and Nonlinear Regimes}

\author{Bhuvnesh Jain}
\affil{Max-Planck-Institut f\"ur Astrophysik,
85740 Garching, Germany}
\affil{bjain@mpa-garching.mpg.de}

\author{Uro\v s Seljak}
\affil{Center For Astrophysics, Harvard University, Cambridge, MA
02138 USA}
\affil{useljak@cfa.harvard.edu}

\def\bi#1{\hbox{\boldmath{$#1$}}}
\begin{abstract}
Weak lensing by large scale structure induces correlated 
ellipticities in the images of distant galaxies. The two-point correlation 
is determined by the matter power spectrum along the line of sight. 
We use the fully nonlinear evolution of the power spectrum to 
compute the predicted ellipticity correlation. 
We present results for different measures of
the second moment for angular scales $\theta \simeq 1'-3^\circ$, 
and for alternative normalizations of the power spectrum, 
in order to explore the best strategy for constraining the cosmological 
parameters. Normalizing to observed cluster abundance the 
rms amplitude of ellipticity within a $15'$ radius 
is $\simeq 0.01\, z_s^{0.6}$, 
almost independent of the cosmological model, with
$z_s$ being the median redshift of background galaxies. 

Nonlinear effects in the evolution of the power spectrum
significantly enhance the ellipticity for $\theta < 10'$ --- for
$\theta\simeq 1'$ the rms ellipticity is $\simeq 0.05$, which is 
nearly twice as large as the linear prediction. This enhancement
means that the signal to noise for the ellipticity is only weakly 
increasing with angle for $2'<\theta<2^\circ$, 
unlike the expectation from linear theory that signal
to noise is strongly peaked on degree scales. 
The scaling with cosmological parameters also changes 
due to nonlinear effects. By measuring the
correlations on small (nonlinear) and large (linear) angular scales,
different cosmological parameters can be independently constrained 
to obtain a model independent estimate of both power 
spectrum amplitude and matter density $\Omega_m$. 
Nonlinear effects also modify the probability distribution of the
ellipticity. Using second order perturbation theory we find that  
over most of the range of interest 
there are significant deviations from a normal distribution. 

\end{abstract}

\section{Introduction}

Mapping the large-scale structure (LSS) of the universe is one of the major
goals of observational cosmology. Traditionally this is performed 
using large surveys of galaxies, either the projected 2-dimensional 
distributions or the 3-dimensional surveys in redshift space. 
The main shortcoming of the
galaxy surveys is that they trace light, while most of the matter 
appears to be dark. One therefore needs to translate the galaxy power 
spectrum into the matter power spectrum and in order to do so one has
to make some assumptions on the nature of galaxy biasing. Even in the simplest 
model this can be achieved only up to an unknown biasing parameter $b$, 
which at present cannot be theoretically estimated.
While sophisticated
N-body and hydro-dynamical simulations will eventually provide some 
answers to this question, at
present the biasing relation between the light and matter remains rather
poorly understood and prevents one from drawing definitive conclusions
on the amplitude and distribution of mass fluctuations in the universe
from galaxy survey data. 

It is clearly important to seek ways to
estimate the large-scale structure that are insensitive to biasing.
Several observable tracers have been proposed 
that probe directly the underlying mass distribution: 
cosmic microwave background (CMB)
anisotropies, gravitational lensing, peculiar velocity flows and 
abundances of massive nonlinear objects.  
This paper focuses on tracing the dark matter with gravitational 
lensing, in particular the effects of weak lensing 
by large scale structure on background galaxies. 
Weak lensing magnifies and shears
the images of distant galaxies. The shear induces an ellipticity in the
image of an intrinsically circular galaxy. Background galaxies
are of course not circular, but by averaging over
the observed ellipticities of a large number of galaxies, the induced
ellipticity can be measured, and related to the mass fluctuations
along the line of sight and to the spatial geometry of the universe. 
Ellipticities of distant background galaxies averaged over several 
arcminute windows are sensitive to the mass power spectrum on scales 
of $1-10 h^{-1}$ Mpc. For a given spectrum of mass fluctuations, it is
sensitive to the cosmological parameters $\Omega_m$ and
$\Omega_\Lambda$. 
Thus while strong lensing which leads to multiple images 
probes non-typical regions of the universe which contain massive halos, 
weak lensing provides a different and a more direct
measure of the mass fluctuations on large scales. 

The first calculations of the shear signal due to weak lensing
that used modern models for the large scale structure power spectrum
were those of \cite{Blandford91,Escude91} 
and \cite{Kaiser92}, based on the pioneering work by Gunn (1967). 
Our work generalizes the results of these authors
to include the effects of nonlinear evolution of the matter
fluctuations for flat as well as open and $\Lambda-$dominated 
cosmologies. Villumsen (1996) has considered some aspects of the 
linear calculation for open models. Very recently, Bernardeau et
al. (1996), Kaiser (1996) and Stebbins (1996) 
have also made the linear calculation
for different cosmologies. Our work extends the results of the above
authors by using the nonlinear power spectrum, which makes a
significant difference on small angular scales of $\theta<10'$. 
For the nonlinear calculation we have used the prescription of 
\cite{Hamilton91} as implemented by \cite{Jain95} and \cite{PD96} 
which provide accurate power spectra for different
models, valid from the linear to the strongly nonlinear regime. 

One can take two different approaches in interpreting a possible
measurement of the shear signal. The first is to work within the
framework of a physical model for the dark matter and background
cosmology. Such a model is best normalized to COBE and 
then its small scale predictions can be compared to 
the observational constraints of e.g. cluster  
or damped Ly-$\alpha$ system abundances, strong lensing statistics 
or peculiar velocity flows,
or in our case the shear amplitude on a given angular scale.  
Another possibility, which is less model dependent, 
is to compare the constraints from different tracers
on the same physical scale. 
This way one can test the gravitational instability 
assumption that both tracers probe the same underlying power spectrum and 
place constraints on the cosmological parameters which scale differently 
with the tracers. 
At the present the CMB data constrain the power spectrum only
on very large scales, where there are no available data from other tracers.
While several current CMB experiments are approaching the 
scales probed by other tracers,  
cosmological parameters such as baryon density, reionization epoch and
Hubble constant become important and complicate the 
power spectrum reconstruction.
The test mentioned above is possible between the cluster abundances
(or strong lensing statistics of multiple images at large separation,
both of which trace essentially the same property) and the 
peculiar velocities.
Unfortunately this method cannot give model independent constraints on 
cosmological parameters, because the two tracers 
scale roughly equally with the matter density $\Omega_m$
(Peebles 1980, White, Efstathiou \& Frenk 1993). 

A similar comparison can also be made between weak lensing observations
and the tracers discussed above. For example, 
the shear amplitude on $15'$ angular scale and
rich cluster abundance both probe power spectrum scales 
around $8 h^{-1}$ Mpc, so the ratio of the two is roughly independent 
of the shape of the power spectrum. One could hope that using such 
comparisons would lead to an estimate of the 
mean density of the universe, because the two tracers scale differently
with it. Unfortunately as shown in this paper this is not the case
so that this is not a promising method of obtaining the mean density. 
Nevertheless, weak lensing can also give important constraints on the power
spectrum both on smaller and larger scales, where other tracers give less
stringent constraints. On the smaller angular scales, our
calculations provide the enhancement in the amplitude and the change
in the scalings with $\sigma_8$ and $\Omega_m$ that arise due to
nonlinear evolution. Once reliable detections of the shear at different
angular scales are available, these results can be used to 
constrain cosmological parameters and the 
matter power spectrum. The principal advantage of working with weak
lensing is that no assumptions about the formation of observable
structures need to be made to connect the mass
power spectrum to the observed shear. The only unknown parameter is
the redshift distribution of source galaxies, and in principle this
can also be determined observationally. 

In \S 2 and the Appendix, the 
formalism for the weak lensing calculation is presented following 
Seljak (1995, 1996), who generalized the work by 
\cite{Blandford91,Escude91} and \cite{Kaiser92} to a non-flat universe and
non-linear regime. The derivation presented here is 
complementary to recent derivations by Bernardeau et al. (1996) 
and Kaiser (1996). 
The linear and nonlinear evolution of the power
spectrum is introduced in \S 2.1, and the resulting dependence on 
cosmological parameters for power law spectra is obtained in \S 2.2. 
In \S 3 realistic CDM-like spectra are used to predict the rms shear.
We provide accurate power law fits for the dependence of the shear on 
the parameters $\theta$, $z_s$, $\sigma_8$ and $\Omega_m$ for
different cosmological models. 
\S 3.1 provides predictions of the shear for COBE and
cluster-abundance normalized spectra for the range of
angular scales which are being probed by current and forthcoming observations. 
The effects of non-gaussianity in the distribution of shear are
considered in \S 4 where we compute the skewness of the
distribution. This is followed by a discussion and conclusions in \S 5.

\section{Theory of weak lensing}

Gravitational lensing shears and magnifies the images of distant
galaxies. The relation of the shear to perturbations in the
gravitational potential along the line of sight is developed in 
the appendix. Here we shall use the formalism presented in the 
appendix to derive expressions for 3 different measures of the second
moment of the shear. 

The observable mean ellipticity of galaxy images can be simply
defined in the approximation that the images are ellipses with
complex eccentricity $\bi{\epsilon}$ given by the axes lengths $b$ and $a$
as: $\bi{\epsilon}=(b^2-a^2)/(b^2+a^2)\ e^{2i\psi}$, where $\psi$ is the
position angle of the major axis. In the limit of weak distortions, the 
eccentricity is the same as ellipticity 
$\bi{\epsilon}\approx (1-b/a)\, e^{2i\psi}$. 
While individual galaxies have 
intrinsic ellipticities which cannot be 
separated from gravitational stretching, we have assumed here that 
averaging over several galaxies leaves only the gravitational
component which can in principle be measured. 
This mean ellipticity $\langle\bi{\epsilon}\rangle$ will be denoted as
$\bi{p}$ and is defined by equation (\ref{shear3}) in the
appendix. We shall use $p$ to denote its amplitude in what follows. 

Information on the ellipticity is contained in the 
trace-free part of the shear tensor, which is obtained by
integrating over all the deflectors between us and the galaxy 
and over the distribution of background galaxies $W(\chi)$
(see appendix),
\begin{eqnarray}
\Phi_{ij} &\equiv & {\partial \delta \theta_i \over \partial \theta_j}
=- 2 \int_0^{\chi_0}
g(\chi)\nabla_i \nabla_j \phi(\chi) d\chi \nonumber \\
g(\chi)&=&r(\chi) \int_\chi^{\chi_0}
{r(\chi' -\chi) \over r(\chi')}W(\chi')d\chi'\ .
\label{shear}
\end{eqnarray}
Here $\phi$ is the gravitational potential and $\chi$ is the radial 
comoving distance with $\chi_0$ being the horizon distance.
We also introduced the comoving angular distance $r(\chi)$ defined
in equation (\ref{rchi}).
The radial distribution of background galaxies is described with
the window $W(\chi')$. 
If all the galaxies are assumed to lie at the same redshift
$z_s$, corresponding to $\chi_S$, 
then $W(\chi')=\delta_D(\chi'-\chi_S)$, and $g(\chi)$ takes
the simple form,
\begin{equation}
g(\chi)=r(\chi)\, \frac{r(\chi_S-\chi)}{r(\chi_S)}\ .
\label{delta}
\end{equation}

There are several two-point statistics that can quantify the induced
ellipticites in galaxy images. 
The easiest to obtain from observations is the rms ellipticity
$\bar{p}(\theta)\equiv 
\langle\bar{\bi{p}}(\theta)\bar{\bi{p}}(\theta)^{*}\rangle^{1/2}$,
where the  overbar indicates the average within 
a circular aperture of radius $\theta$. Equation (\ref{shear}) can be
used to express $\bar{p}(\theta)$ in terms of 
the power spectrum of density perturbations $P_{\delta}(k,\chi)$. 
Following the derivation presented in the appendix
we obtain for the ellipticity variance
\begin{eqnarray}
\bar{p}^2(\theta) &=&9(2\pi)^2\Omega_m^2H_0^4\theta^{-2}\int_0^{\chi_0} 
\left({g \over ra}\right)^2d\chi\int
P_\delta(l/r\theta,\chi)W_2^2(l)ldl \nonumber \\
&=&36 \pi^2 \ \Omega_m^2
\int_0^{\infty}\ k dk\int_0^{\chi_0}\ a^{-2}(\chi)\ P_{\delta}(k,\chi)
\ g^2(\chi)
W_2^2[kr(\chi)\theta]d\chi,
\label{p}
\end{eqnarray}
where $W_2(x)=2J_1(x)/x$ with $J_1(x)$ being the Bessel function of
first order and $a$ is the expansion factor normalized to unity today.
A related quantity is the two-point polarization correlation function
$C_{pp}(\theta)=\langle\bi{\epsilon}(0)\,
\bi{\epsilon}^*(\theta)\rangle$, 
which is given by a similar expression to the one above, replacing
the term $W_2^2$ with $J_0$:
\begin{equation}
C_{pp}(\theta)=36 \pi^2\  \Omega_m^2
\int_0^{\infty}\ k dk\int_0^{\chi_0}\, a^{-2}(\chi)\ P_{\delta}(k,\chi)
\ g^2(\chi)\ J_0[kr(\chi)\theta]\, d\chi.
\label{Cpp}
\end{equation}
The Fourier transform of $C_{pp}(\theta)$ is the angular 
ellipticity power spectrum $P(l)$, which is often the 
optimal statistic to use (Kaiser 1992). Here we use 
the quantity $\sigma^2(l)=
2\pi l^2P(l)$, which
gives the contribution to the variance per log-interval
in $l$ and is given by
\begin{equation}
\sigma^2(l)=36 \pi^2\ \Omega_m^2\ l^2\, 
\int_0^{\chi_0}\ {g^2(\chi) \over r^2(\chi)}\ 
a^{-2}(\chi)\ P_\delta\left(k={l \over r(\chi)},\chi\right)d\chi.
\label{ps}
\end{equation}
This gives an estimate of correlations at an angle $\theta=1/l$.

No assumption on the matter power spectrum has been made in the above 
expressions and one can use them both in the linear and nonlinear regime. The
nonlinear spectrum is in general a non-separable function of
wavenumber and redshift, thus leading to a complicated coupling 
of the dependences on the distance factors and the growth of perturbations. 
It also leads to a nonlinear dependence of
the predicted ellipticity correlations 
on the shape and amplitude of the initial power
spectrum. Thus the amplitude of the ellipticity correlations as well
as their dependence on cosmological parameters can change in the
nonlinear regime (i.e., on small angular scales). 
In the rest of the paper we shall explore in detail
the outcome of these effects of nonlinear evolution. 

\subsection{Evolution of the power spectrum}

If the gravitational potential changes in time its power 
spectrum will depend on the radial distance $\chi$. 
In linear theory 
this dependence is independent of $k$ and can be written in terms of 
the potential growth factor $F(\chi)$ as
$P_\phi(k,\chi)=F^2(\chi)P_\phi(k)$, which gives for the density 
power spectrum $P_\delta(k,\chi)/a^2(\chi) =
F^2(\chi)P_\delta(k)=[(D_+/a)^2](\chi)P_\delta(k)$, where  $D_+(\chi)$ is  
the linear growth factor for the density.  
This can be approximated as (Lahav et al. 1991)
\begin{eqnarray}
F(\chi) = 2.5\, \Omega_m\, a^{-1}\, (xf+1.5\Omega_m a^{-1}+
\Omega_K)^{-1} \nonumber\\
x=1+\Omega_m(a^{-1}-1)+\Omega_\Lambda(a^2-1)\, \, 
;\, f=\left({\Omega_m \over ax} \right)^{0.6}.
\label{f}
\end{eqnarray}
We have ignored the weak $\Omega_\Lambda$ dependence of the logarithmic 
growth factor. 
For $a=1$, this expression simplifies to
\begin{equation}
F(\chi=0)=2.5\Omega_m\, (1+\Omega_m^{0.6}+0.5\, 
\Omega_m-\Omega_\Lambda)^{-1}\, .
\label{f0}
\end{equation}
For a flat $\Omega_m=1$ model the gravitational potential does
not change in time in the linear regime and the region where $g(\chi)$ peaks
dominates the radial integral in equations (\ref{p}-\ref{ps}). Typically
this is at half the mean comoving distance to the galaxies, so if
background galaxies lie at 
$z=1$ typical deflectors lie at 
$z \approx 0.3$. In low $\Omega_m$ models 
the gravitational potential increases with $\chi$, differently for open and 
for cosmological constant models. 
In the small $z$ limit the growth factor
only depends on $\Omega_m$ (Villumsen 1996). 

On scales where $\Delta^2(k, \chi)=4\pi k^3P_\delta(k, \chi)$ 
approaches or exceeds
unity nonlinear evolution of the power spectrum becomes important.
This is particularly important if one is discussing 
low $\Omega_m$ models, where both cluster abundances and peculiar 
velocity normalizations give higher density normalization
on scales of interest.
In the quasilinear regime the nonlinear spectrum can be computed
using perturbation theory. An alternative semi-analytic approach
which is accurate from the linear to the strongly nonlinear regime
(up to density contrasts $\sim 10^3$) was proposed by Hamilton et
al. (1991) and further developed in subsequent work (e.g., Nityananda
\& Padmanabhan 1994; Peacock \& Dodds 1996; Jain, Mo \& White
1995; Padmanabhan et al. 1996). 
This prescription involves mapping the nonlinear spectrum at a wavenumber
$k$ to the linear spectrum at a unique wavenumber $k_L<k$ which is
given by a spherical collapse model. This leads to a mapping of the 
form $\Delta^2(k, \chi)=G\left[\Delta_L^2(k_L, \chi)\right]$, where 
the dimensionless power $\Delta^2$ is defined above, and $G$ is
a function which varies in a simple way for different initial spectra. 
The linear wavenumber $k_L$ is given by $k_L=k\, [1+\Delta^2(k,
\chi)]^{-1/3}$. The functional form of the mapping is thus specified
by the function $G$, which has
been calibrated using high resolution N-body simulations and is
accurate for a wide range of initial spectra. 

For the predictions of the rms ellipticity in \S 3 
we shall use the fitting formulae of
\cite{Jain95} for $\Omega_m=1$ and of \cite{PD96} for the open
and $\Lambda-$models to describe the nonlinear power spectrum as 
a function of wavenumber and redshift. We refer the reader to 
the above references for details of the formulae and their
implementation. While the formulae can differ by a few tens of percent for 
certain spectra, we find that our results for the ellipticity 
are not affected by more than a few percent as the ellipticity integrates
the power spectrum over a range of wavenumbers and redshifts. 
In particular the results for the CDM-like models shown in 
Figures 4-9 would be completely unaffected by our choice of fitting
formulae. 

\subsection{Dependence on $z_s$, $\sigma_8$ and $\Omega_m$ for
power law spectra}

In the next section we shall use the nonlinear evolution of CDM-like
spectra to predict the rms ellipticity signal. Here we consider the
simplified case of power law initial spectra 
$P_\delta(k)=A k^n$ to compute
the scalings of $C_{pp}(\theta)$ and $\bar{p}^2(\theta)$ with source
redshift $z_s$, and the cosmological model 
parameters $\sigma_8$ and $\Omega_m$. The parameter $\sigma_8$ is
the rms fluctuation in the mass on scales of $8 h^{-1}$Mpc, and serves
to normalize the power spectrum. We use the linear
evolution of the power spectrum for which some of the scalings
can be obtained analytically. We also consider the qualitative
modification due to nonlinear evolution by using the stable clustering
regime which gives the maximal effect of nonlinearities. 
The length scales that contribute to the
ellipticity are of order $8 h^{-1}$ Mpc for $\theta \simeq 15'$. On these
scales most realistic spectra have a spectral index $n\simeq
-1$. On smaller angular scales, the contribution to the ellipticity comes
from higher wavenumbers which sample a steeper part of the spectrum
(lower $n$). We therefore choose power law spectra with
$n=-1$ and $n=-2$ for comparison with realistic spectra on angular
scales of interest. 

For a power law spectrum the $k-$integral for $C_{pp}$ is 
analytic, and gives:
\begin{equation}
C_{pp}(\theta)=36 \pi^2 \, \Omega_m^2\, A\ 
\frac{2^{n+1} \Gamma(1+n/2)}{\theta^{2+n}\, \Gamma(-n/2)} 
\int_0^{\chi_0} d\chi \, g^2(\chi)\,r(\chi)^{-2-n} F^2(\chi) , 
\label{powerlaw}
\end{equation}
valid for $-2<n<-1/2$. The constant $A\propto \sigma_8^2$ 
gives the normalization of the power spectrum. 
Aside from the numerical prefactors, the result for $\bar{p}^2(\theta)$
is the same and involves the same $\chi$ integral. 

For simplicity we shall take all the source galaxies to be at the
same redshift -- this is a good approximation as the result is
insensitive to the source distribution for a given median source
redshift. The window function $g(\chi)$ then takes the form given
in equation (\ref{delta}). 
For $\Omega_m=1$, $r(\chi)=\chi$, and
the $\chi$ integral is analytic as well giving the following 
power law dependence: 
\begin{equation}
C_{pp}\, ,\, \bar{p}^2 \, \propto\, \sigma_8^2\ \chi_S^{1-n}\ \theta^{-2-n}\, . 
\label{powerlawflat}
\end{equation}
Using $\chi=2H_0^{-1} (1-a^{1/2})$, the dependence of $C_{pp}$
on the source redshift $z_s$ can be obtained from the above equation. 

For $\Omega_m<1$ and $\Omega_\Lambda > 0$, the result for $C_{pp}$ is not
analytic except in the limit of $z\ll 1$ which is not of interest 
for realistic situations. It is simple to understand qualitatively
the two physical effects that enter. 
$(i)$ For a given redshift, the distance $\chi$
increases as $\Omega_m$ decreases and $\Omega_\Lambda$ increases. Therefore
there is more path length in the line of sight integral as $\chi_S$ 
is larger for a given $z_s$. The factor $g(\chi)$ involves factors of
the comoving angular diameter distance 
$r(\chi), r(\chi_S)$ and $r(\chi_S-\chi)$; it 
also increases with decreasing $\Omega_m$ and increasing
$\Omega_\Lambda$.  
(ii) The second factor is the linear growth factor $F(\chi)$. 
Growth of structure slows down at low redshift in an open universe, 
and to a lesser extent in a $\Lambda-$ dominated universe. Therefore,
at a given redshift the linear growth factor, normalized to unity today, 
increases as $\Omega_m$ decreases, and for a given $\Omega_m$ it
decreases as $\Omega_\Lambda$ increases. 

\begin{figure}[p]
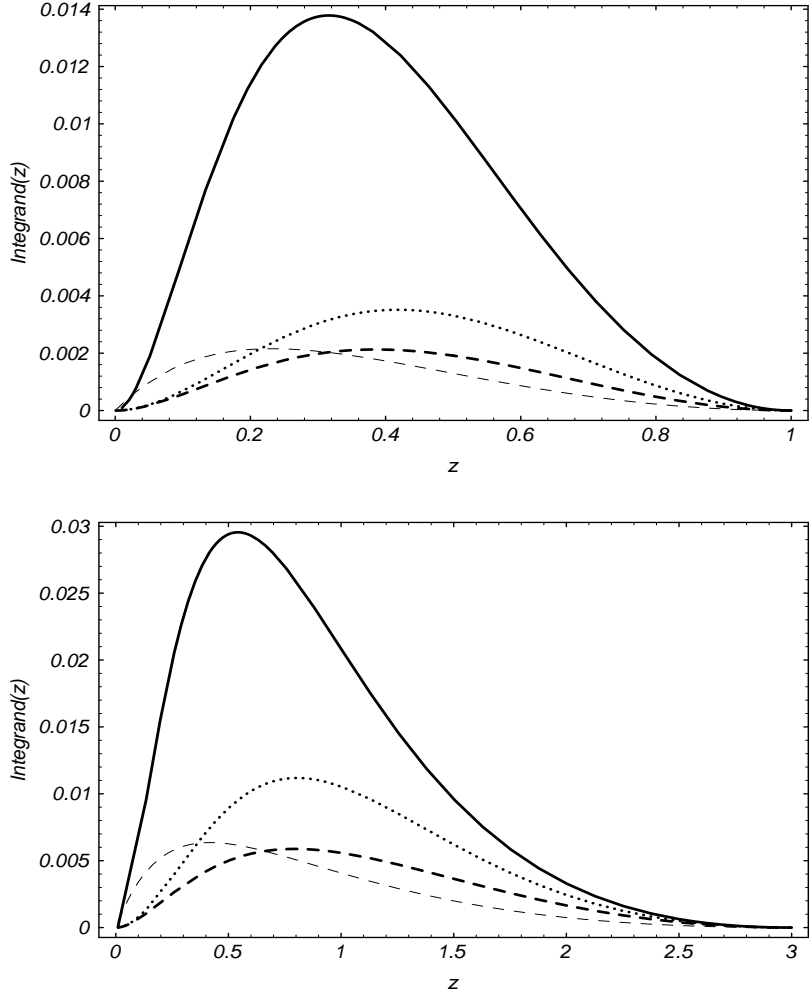

\vspace*{11.3cm}
\caption{The dependence of the integrand for $C_{pp}$ on redshift.
The source galaxies are assumed to be at $z_s=1$ (upper panel), 
and $z_s=3$ (lower panel), and the matter
power spectrum is a power law with slope $n=-2$. The solid curve is 
for  $\Omega_m=1$, the dashed one for $\Omega_m=0.3, \,
\Omega_{\Lambda}=0$, and the
dotted one for $\Omega_m=0.3,\, \Omega_\Lambda=0.7$. 
The figure shows that compared to the Einstein-de Sitter case, 
the amplitude decreases for the $\Lambda$ model and is lowest 
for the open model. The peak of the
integrand for $z_s=1$ lies at about $z=0.3$ for the $\Omega_m=1$ case,
and shifts to higher $z$ for the open and $\Lambda$ models. 
The light, dashed curve is for an $n=-1$ spectrum, with 
$\Omega_m=0.3$. It peaks at lower $z$ compared to the heavy, dashed 
curve as the spectrum has more small scale power.
}
\includegraphics{fig1a.ps}
\includegraphics{fig1b.ps}
\label{fig1}
\end{figure}

The effect of both the distance and growth factors is to increase
the contribution from the $\chi$ integral in equation (\ref{powerlaw}) 
relative to the Einstein-de Sitter case. 
The net contribution increases as $\Omega_m$ decreases, 
and for a given $\Omega_m$, as $\Omega_\Lambda$ increases. In the 
case of $\Lambda$, the distance factors dominate over the linear
growth factor leading to a net increase over the case with the 
same $\Omega_m$. However the enhancement due to the integral over $\chi$
has to contend with the factor
of $\Omega_m^2$ outside the integral which is considerably larger. 
The net result is shown in figure \ref{fig1} for source galaxies
at $z_s=1$ (upper panel) and $z_s=3$ (lower panel). The y-axis shows
$\Omega_m^2$ times the $\chi$-integrand of equation
(\ref{powerlaw}). Thus $C_{pp}$ or $\bar{p}^2$ is proportional to 
the area under the curves shown, aside from numerical factors 
which depend on the shape of the spectrum. The figure shows that
the peak contribution for a reasonable choice of $z_s$ comes from
the range $z \simeq 0.3-0.6$. 

The scaling of $\bar{p}(\theta)$ (the same as that of the square root 
of $C_{pp}$) with 
$\Omega_m$ is dependent both on $z_s$ and the shape of the power
spectrum. For $\Omega_\Lambda=0$ the dependence is well fit by the
power law, 
\begin{equation}
z_s=1\ :\ \bar{p}(\theta) \, \propto\, \sigma_8 \, \Omega_m^{0.85} \, .
\label{scaling1}
\end{equation}
For $z_s\simeq 1$ and $-2\lsim n \lsim -1$, the above equation
is quite accurate. For $z_s=3$ the scaling depends on the 
spectrum, and can be approximated as,
\begin{equation}
z_s=3\ :\ \bar{p}(\theta) \, \propto\, \sigma_8 \, \Omega_m^{0.75}\ \
{\rm for}\ \ n=-1\ \ 
;\ \  \bar{p}(\theta) \, \propto\, \sigma_8 \, \Omega_m^{0.6}\ \
 {\rm for}\ \ n=-2\, .
\label{scaling2}
\end{equation}
The result for $n=-2$ is valid provided a low-$k$ cutoff is imposed
to keep the integral finite. 
As expected, at higher source redshifts the enhancement due to the
distance and growth factors is larger and leads to a weaker
$\Omega_m$ dependence. The same occurs as $n$ decreases, as then
there is more power on large scales, and therefore more weight to the
high-$z$ part of the integral. The results of equations (\ref{scaling1})
and (\ref{scaling2}) are valid for $0.2 \lsim \Omega_m \lsim 1$, and
are intended to cover the range of source redshift and spectra spanned
by realistic models. 

A simple way to understand the dependence on $\Omega_m$ follows from
the fact that most of the lensing contribution comes from about half
the distance to the source galaxies. Thus the $\Omega_m$
dependence of equations (\ref{scaling1}) and (\ref{scaling2}) can be 
compared to that of $\Omega_m$ at the redshift $z_{1/2}$, corresponding 
to $\chi_S/2$. $\Omega_m(z_{1/2})$ is well approximated by 
$\Omega_m^{0.8}$ for $z_s=1$ and by $\Omega_m^{0.6}$ for $z_s=3$ 
if $n=-2$. For a shallower slope there is more power on small scales, 
which are closer for a given angle.
Hence the results are closer to those of equations (\ref{scaling1})
and (\ref{scaling2}) for $n=-1$ if one uses $\chi_S/3$ (it makes
very little difference if one uses $r(\chi)$ instead of $\chi$ for
this purpose). 
Thus as a first approximation the $\Omega_m$ dependence can be obtained
by replacing the $\Omega_m$ term in $C_{pp}$ for the flat case, by 
the value of $\Omega_m$ at $1/3$ to $1/2$ the 
distance to the source galaxies. 

\begin{figure}[t]
\vspace*{9.3cm}
\caption{The transverse distance at $z=0.3$ (lower set of curves)
and $z=0.5$ (upper curves) is shown as a function of the angle. 
The three curves at each redshift are as in figure 1. The two 
redshifts are chosen to span the range of redshifts that provide
the peak contribution for source galaxies at $z_s\sim 1-2$. 
}
\includegraphics{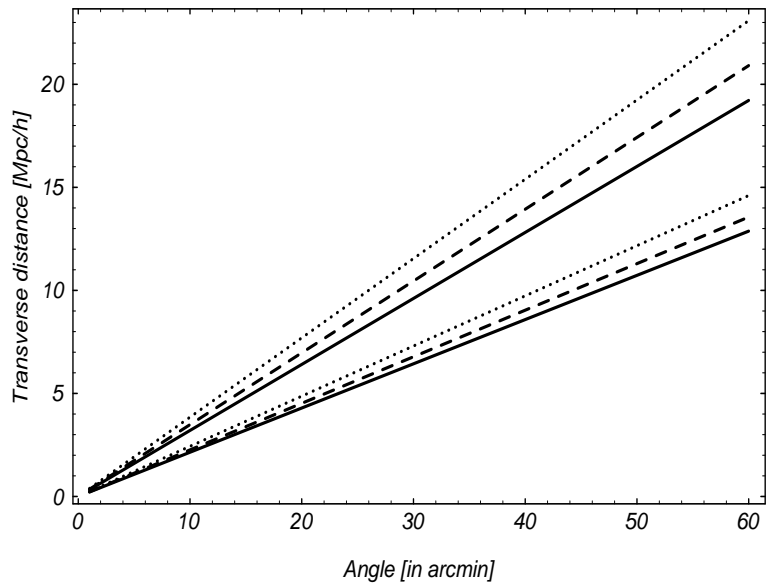}
\label{dist}
\end{figure}

The above scalings are derived using linear evolution of the power
spectrum. On smaller angular scales nonlinear evolution could play
an important role --- in the next section we compute the change in
the above scalings due to nonlinear effects. 
It is complicated to estimate the angular
scale where nonlinear effects are important because each angle
involves the projection of a range of length scales at different
redshifts. The answer will therefore depend on the slope of the 
power spectrum. A rough estimate of the length scales that make the
dominant contribution to $\bar{p}(\theta)$ is given by figure
\ref{dist}. This figure shows the transverse distance for a given $\theta$ 
at the redshifts estimated from figures \ref{fig1}a and \ref{fig1}b
as providing the dominant contribution to $\bar{p}$ for source galaxies
at $z_s\sim 1-2$. For $\theta<10'$ this length scale is less than
$5 h^{-1}$Mpc, which is the correlation length of galaxies and
therefore a good demarcation of the nonlinear regime. 
We may thus expect that the linear results for $\bar{p}$ will 
be modified for $\theta<10'$. 
  
An analytic estimate of the change in the above scalings due to
strongly nonlinear effects can be made using the stable clustering
solution. If one assumes that the lensing is dominated by clustering
on small scales (valid in the limit of small angles and low source
redshift), which stabilized at high redshift, then the dimensionless
power is a function of the physical wavenumber times $a^3$. This 
property extends to the open models as well, because at high enough
redshift, $\Omega_m(z) \simeq 1$. It can then be shown that, if the
linear spectrum is normalized to the same $\sigma_8$ today, then
the enhancement factor of the nonlinear spectrum relative to the linear 
one is proportional to 
$F(\chi)^{-3}$. This follows from a reversed version of the
argument of Peacock \& Dodds (1996), 
which was made for spectra normalized to the same initial value. 
A particularly simple example is provided by the $n=-2$ spectrum, 
for which the nonlinear slope and growth in the Einstein-de Sitter
case are the same as in the linear regime. 
Hence the nonlinear spectrum is given at all $k$ by a constant
enhancement factor relative to the linear spectrum. 

For $0.3\lsim \Omega_m\lsim 1$ one can approximate
$F(\chi)\simeq \Omega_m^{0.5}$ for $z_s\simeq 1$. Therefore 
the $\Omega_m$ dependence in $\bar{p}^2(\theta)$ due to strongly
nonlinear evolution of the power spectrum is $F(\chi)^{-3}\sim
\Omega_m^{-1.5}$. In this regime then, 
$\bar{p}^2 \propto \Omega_m^{0.5}$ times the contribution of the distance
factors, which further lowers the power of $\Omega_m$.
This estimate is meant
to provide an upper bound on the enhancement for open models due to
the nonlinear contribution. For realistic redshifts the contribution
from such small scales does not dominate, and the exact, numerical 
estimates of the nonlinear spectrum must be integrated over $k$ and 
$\chi$. Still, it is clear that the $\Omega_m$ dependence of
weak lensing amplitude will be significantly weaker than the linear 
relation predicted from the scalings in previous subsection.

\section{Predictions for rms ellipticity for CDM-like spectra}

The detailed predictions for the rms ellipticity depend on the
shape and amplitude of the power spectrum, the distribution of source
galaxies, and on the cosmological parameters $\Omega_m$ and
$\Omega_\Lambda$. In the following subsection we shall consider
two alternative normalizations of realistic CDM-like power spectra. 
A useful, model independent description of the scalings can also
be obtained by fitting the dependence on the different parameters
to power laws. This provides a reasonably accurate approximation
to the result, and shows the qualitative behavior more clearly. 

Table 1 provides such power law fits to the scalings of $\bar{p}$
with $\theta$, $z_s$, $\sigma_8$, and $\Omega_m$ for Einstein-de
Sitter, open and $\Lambda-$dominated cosmologies. The three
specific models we have chosen are: 
Einstein-de Sitter ($\Omega_m=1$, $\Omega_\Lambda=0$), Open
($\Omega_m=0.3$, $\Omega_\Lambda=0$), 
and $\Lambda-$dominated ($\Omega_m=0.3,
\Omega_\Lambda=0.7$). The latter two models are representative of 
the class of open and flat $\Lambda-$dominated cosmologies as the
results do not change significantly for $\Omega_m$ in the range
$0.2<\Omega_m<0.5$. The matter power spectrum used in the calculations
is the nonlinear $\Gamma=0.25$ CDM spectrum, where $\Gamma$ is
the shape parameter defined in Bardeen et al. (1986) and roughly
corresponds to $\Omega_mh$ in CDM models. The scalings given in table 1 
are not strongly sensitive to the shape of the power spectrum as shown in
figure \ref{shape} and discussed in the next subsection. 

\begin{table}[t]
\centering
\caption{Scaling of the polarization $\bar{p}(\theta)$ with $\theta$,
$z_s$ and with cosmological model parameters}
\bigskip
\begin{tabular}{|c|c|c|c|c|}   \hline

$\theta$   &  $z_s$    &   $\sigma_8$  &  $\Omega_m$ & $\Omega_\Lambda$\\ 
\hline
 $\theta^{-0.37}$  & $1$     &  $1$ & 1 & 0 \\
 $\theta^{-0.47}$   & $1$     & $1$ & 0.3 & 0\\ 
 $\theta^{-0.42}$   & $1$     & $1$ & 0.3 & 0.7\\ \hline
  $2'-5'$   &   $z_s^{0.6-0.57}$  &    $1$ &   1& 0\\
  $15'$   &   $z_s^{0.56}$  &    $1$  &   1& 0\\
  $2'-5'$   &   $z_s^{0.66-65}$  &    $1$ &   0.3& 0\\
  $15'$   &   $z_s^{0.67}$  &    $1$  & 0.3& 0\\ 
  $2'-5'$   &   $z_s^{0.77-74}$  &    $1$ &   0.3& 0.7\\
  $15'$   &   $z_s^{0.74}$  &    $1$  & 0.3& 0.7\\ \hline
  $2'-5'$   &   $1$  &    $\sigma_8^{1.25-1.20}$ &   1& 0\\
  $2'-5'$   &   $1$  &    $\sigma_8^{1.38-1.34}$ &   0.3& 0\\ 
  $2'-5'$   &   $1$  &    $\sigma_8^{1.29-1.27}$ &   0.3& 0.7\\ \hline
  $2'-5'$   &   $1$  &    $1$ & $\Omega_m^{0.66-0.75}$& 0\\
  $15'$     & $1$  &    $1$ & $\Omega_m^{0.81}$& 0\\
  $2'-5'$   &   $1$  &    $1$ & $\Omega_m^{0.60-0.65}$& 0.7\\
  $15'$     & $1$  &    $1$ & $\Omega_m^{0.68}$& 0.7\\
  $2'-5'$     & $3$  &    $1$ & $\Omega_m^{0.60-0.65}$& 0\\
  $15'$     & $3$  &    $1$ & $\Omega_m^{0.80}$& 0\\
\hline 
\end{tabular}
\label{table}
\end{table} 
 
\begin{figure}[t]
\vspace*{8.3cm}
\caption{The ratio of $\bar{p}(\theta)$ computed using the nonlinear
spectrum to that with the linear $\Gamma=0.25$ CDM spectrum is shown. 
The normalization is $\sigma_8=1$, and $z_s=1$. 
The solid curve is for $\Omega_m=1$, the dashed curve for 
$\Omega_m=0.3$ and the dotted curve for $\Omega_m=0.3$,
$\Omega_\Lambda=0.7$. The results show the significant enhancement in
$\bar{p}(\theta)$ due to nonlinear evolution for $\theta<10'$. 
}
\includegraphics{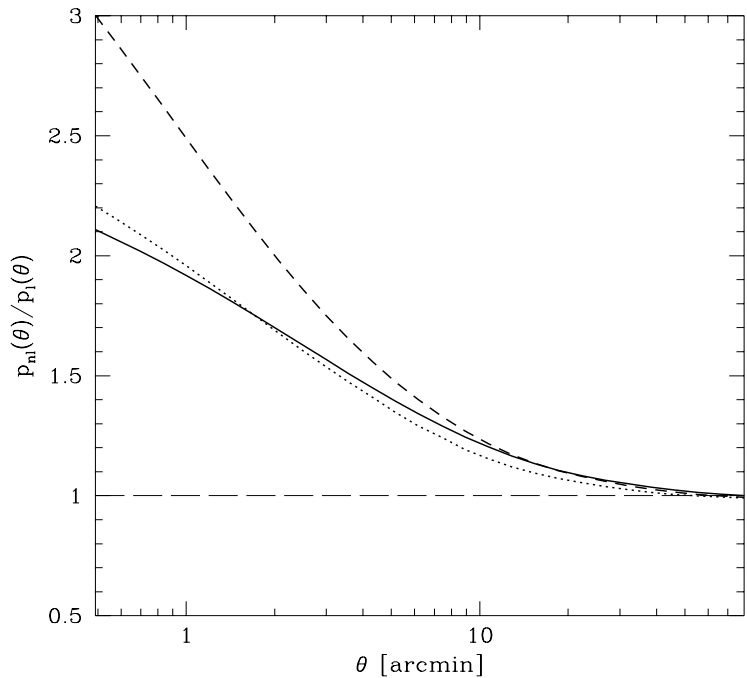}
\label{ratio}
\end{figure}

On scales $\theta<10'$ nonlinear evolution makes a significant difference
to all the scalings. Figure \ref{ratio} shows the ratio of $\bar{p}(\theta)$
computed using the nonlinear/linear power spectrum 
for the above three models. The ratio is significantly larger than
unity for $\theta<10'$, and reaches a factor of two for $\theta\sim
1'$. It is the largest for the open model, where the nonlinear effects
are more important because of a slower linear growth of perturbations. 

Since the scalings differ in the linear and nonlinear regimes, table 1
gives power law fits for three different angles, $\theta=2', 5',
15'$, to represent the nonlinear, weakly nonlinear and linear regimes,
respectively. These give $\bar{p}(\theta)$ to good accuracy
over most of the range: $1'< \theta< 30'$; 
$0.5<z_s<3$; $0.5<\sigma_8<2$; $0.2<\Omega_m<1$. 
The results can be approximated by the following equations for the three 
cosmological models given above. The angle $\theta$ below is in arcminutes,
and where the scalings differ on $2'$ and $15'$, the latter scaling
is given in brackets in italics. Since the scaling with $z_s$ is
nearly the same for the two values of $\theta$, we have used an intermediate
value in the following equations. 
\begin{eqnarray}
\bar{p}\left[2'({\it 15'})\right]\ &=&\ 0.08\ \ \theta^{-0.37} \ z_s^{0.58}\ 
		\sigma_8^{1.25 ({\it 1})}\ \Omega_m^{0.66 ({\it
		0.81})} \ \ \ \ \ \ \ \ \ :\ \Omega_m=1.0\, 
\nonumber \\
\bar{p}\left[2'({\it 15'})\right]\ &=&\ 0.04\ \ \theta^{-0.47}\ 
		 	z_s^{0.66}\ \sigma_8^{1.38 ({\it 1})}\ 
			(\Omega_m/0.3)^{0.66
		 	({\it 0.81})} \ :\ \Omega_m=0.3\, 
\nonumber \\
\bar{p}\left[2'({\it 15'})\right]\ &=&\ 0.04\ \ \theta^{-0.42}\ 
		 	z_s^{0.76}\ \sigma_8^{1.29 ({\it 1})}\ 
			(\Omega_m/0.3)^{0.60
		 	({\it 0.68})} \ :\  \Omega_\Lambda=0.7\, .
\label{scalings-lambda}\end{eqnarray}

These power law fits can be used to obtain predictions for the rms
ellipticity for any desired choice of cosmological parameters. 
For $\theta=15'$, the results are consistent with the linear theory
estimates for power law spectra (with $n$ between $-1$ and $-2$) 
obtained in the previous section. 
The $\bar{p}(\theta)$ curve is enhanced at small angles due to
nonlinear evolution. This enhancement causes 
the dependence on $\theta$ to be much closer to
a power law over the range $1'<\theta<30'$ than the linear prediction,
which has much greater curvature due to the curvature of the 
power spectrum. 
This power law is steeper for the open and $\Lambda$ models due to
stronger nonlinear evolution. 

On $\theta\simeq 2'$, the dependence
on $\sigma_8$ is significantly stronger than linear, especially for
the open model. Nonlinear evolution also weakens the dependence on
$\Omega_m$ as anticipated in the previous section. 
Combining the scaling with $\sigma_8$ and $\Omega_m$ (relevant for
comparison with other tracers), $\bar{p}(\theta)$ for $\theta\simeq 2'$ 
scales as $\sigma_8^{1.25}\, \Omega_m^{0.66}$, whereas on large 
scales of $\theta>10'$, it scales as $\sigma_8\, \Omega_m^{0.8}$.
Thus on small scales  $\bar{p}(\theta)$ measures
$\sigma_8\Omega_m^{0.5}$ (the power of $\Omega_m$ is even lower for 
$\theta<2'$ or $\Omega_m<0.5$), whereas on large scales it measures
$\sigma_8\, \Omega_m^{0.8}$. 

Using the above differences a comparison of $\bar{p}(\theta)$ at small and 
large angular scales can constrain $\sigma_8$ and $\Omega_m$
separately. Thus nonlinear evolution, at the expense
of a more complex dependence on various parameters, can help break
the degeneracy between these parameters. 
To distinguish an open from a flat
$\Lambda-$model with the same value of $\Omega_m$ 
one needs to use the difference in the scaling with $z_s$, 
as the other scalings are very similar. 
The $\Lambda-$model predicts a faster increase with $z_s$ due to
a faster increase of the distance factors with redshift 
than the open case. Once it
becomes feasible to get estimates of the redshifts of source galaxies,
this difference can become a useful test of $\Lambda$ as it is 
nearly independent of the shape and normalization of the power
spectrum. 

\subsection{Results for cluster-abundance and COBE normalized spectra}

In choosing the most realistic power spectrum 
to compute the polarization signal, the approach closest to other
tracers of the mass density field would be to use the abundance of
galaxy clusters or the peculiar velocity field of galaxies to 
estimate the amplitude $\sigma_8$, and to then use the galaxy 
power spectrum to constrain the shape 
(Peacock \& Dodds 1996; Peacock 1996). 
The merit of this approach is that it agrees with the observational
constraints on the scales which are most relevant for weak lensing 
observations.
An alternative
is to work with physical models for the dark matter and cosmology
which give the shape of the power spectrum, and use COBE-normalization
to fix the spectrum on very large scales and thus its overall
amplitude. The merit of this approach is that for a given 
model both the spectrum of perturbations and their evolution is 
fully determined.
However, the final spectrum may or may not agree with the present day
observations (which interpretation is still somewhat uncertain).
We shall present results using both approaches. 

The first of the two reliable tracers of
mass fluctuations on large scales is the abundance 
of large galaxy clusters with a typical mass of $10^{15}M_\odot$,
which corresponds to a linear scale of about $8 h^{-1}$Mpc. These
objects are very rare and according to the Press-Schechter formalism
(Press \& Schechter 1974) 
their number density depends exponentially on the amplitude
of mass fluctuation on the corresponding linear scale. This allows 
an accurate estimate of the amplitude, which typically gives 
$\sigma_8=0.5-0.6$ (White, Efstathiou \& Frenk 1993, 
Viana \& Liddle 1995, Eke, Cole \& Frenk 1996, Pen 1996) 
in an $\Omega_m=1$ universe, 
almost independent of the shape of the power spectrum. 
At present this gives the best constraint on the linear density power spectrum,
with an error of about 10-20\%.
The dependence on $\Omega_m$ is approximately
$\sigma_8 \propto \Omega_m^{-(0.5-0.6)}$, and differs only slightly for open
and spatially flat models. The scaling with 
$\Omega_m$ and $\Omega_\Lambda$ is still somewhat uncertain as it requires 
calibration with large N-body simulations.

Peculiar velocities are also tracers of underlying mass fluctuations. In the 
linear regime there is a simple relation between velocity and density fields 
(Peebles 1980):$\delta(r)=-(H_0f)^{-1}{\bi \nabla \cdot \vec v(r)}$,
where $f\simeq \Omega_m^{0.6}$. 
The $\Omega_m$ dependence on mass fluctuation amplitude from 
velocity data is $\sigma_8 \propto \Omega_m^{-0.6}$, which is almost the same
as for the cluster abundance in a flat universe and  
only slightly less so in an open universe. Comparison of the 
two constraints therefore cannot significantly constrain $\Omega_m$.
Given a set of velocity measurements 
one can reconstruct the particular combination $f\delta$ and estimate its
power spectrum. Recent comparisons between Mark III Tully-Fisher  
catalogs and IRAS survey tend to favor 
$\beta_{I}=\Omega^{0.6}/b_I \sim 0.5$ 
(\cite{wsdk}, \cite{dnw}), 
where $b_I$ is the linear bias parameter for IRAS galaxies. This 
combined with $\sigma_{8,I} \sim 0.7$ leads to $\sigma_8\Omega^{0.6}=0.35$, 
smaller than the value derived from the cluster abundance data.
This would therefore reduce the weak lensing predictions presented in 
this paper. However, analysis of Mark III catalog with POTENT 
gives a higher value of $\sigma_8\Omega^{0.6}$ 
(Kolatt \& Dekel 1996), so it seems prudent to adopt the cluster 
abundance normalization until the discrepancy in the velocity data 
is resolved.

Based on the above discussion, a useful functional form for the 
density power spectrum is given by a   
CDM type transfer function with two free parameters, the 
amplitude $\sigma_8 \approx 0.6 \Omega_m^{-0.6}$ 
and the shape parameter $\Gamma \approx 0.25$ 
(Peacock \& Dodds 1996). This is the model that will be adopted in 
computing predictions of ellipticity polarizations for
cluster-abundance normalization. 
We shall use the fitting formulae in equations (47-49) 
of \cite{Viana95} to compute $\sigma_8$ as a function of $\Omega_m$ -- 
their normalization is very close to the  
results of \cite{WEF93} and Pen (1996), though a bit higher than that of 
\cite{Eke96}. 

\begin{figure}[t]
\vspace*{9.3 cm}
\caption{The logarithmic contribution from wavenumbers $k$ to 
$C_{pp}$ (dashed) and $\sigma^2(1/\theta)$ (dot-dashed) 
at $\theta=15'$, $z=1$ is shown. It is compared to that of $\sigma^2_8$
(solid), the variance in density on $8 h^{-1}$Mpc scale. 
The dependence of $\bar{p}^2$ is nearly the same as that
of $C_{pp}$ up to $k\simeq 0.5 h$Mpc$^{-1}$, after which it remains
positive while $C_{pp}$ goes negative. 
}
\includegraphics{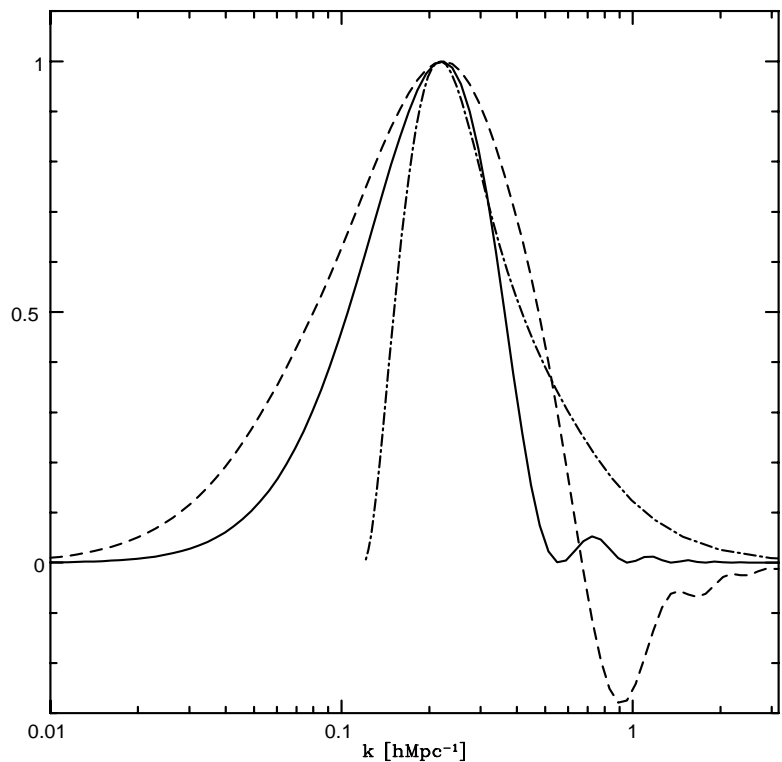}
\label{fig11}
\end{figure}

Before proceeding with the results for $\bar{p}(\theta)$ we consider
the contribution to the integral from different ranges in $k$. 
Figure \ref{fig11} compares the contribution per logarithmic 
interval in wavenumber to $\bar{p} 
^2$ and $\sigma^2(l)$ at $\theta=15'$, $z_s=1$
with the logarithmic contribution to $\sigma_8^2$, which is also
an integral of the power spectrum over $k$ with a real space top-hat filter.
All the distributions 
are normalized so that they peak at unity. One can see that  
the distribution of $\bar{p}^2$ is broader and that of $\sigma^2(l)$ 
narrower than the corresponding $\sigma_8^2$
distribution, but in general they sample very much the same scales. 
Thus the optimal statistic to use in order to compare the
normalization $\sigma_8$ with other tracers 
is $\sigma^2(l)$, $\bar{p}^2(\theta)$ at $1/l$, $\theta \simeq 15'$. 
The comparison can then 
be made nearly independent of the shape of the power spectrum. 

We need to choose a redshift distribution for the source galaxies in 
order to compute $\bar{p}(\theta)$. At present this is spectroscopically 
known only for galaxies with magnitude in I below 23 (Lilly et al. 1996; 
Cowie et al. 1996), 
which have median redshift $z \sim 0.5$, while typical lensing observations 
reach several magnitudes deeper (e.g. Mould et al. 1994). 
HST observations of Abell 2218 coupled with cluster mass reconstruction 
indicate that the median redshift increases from 
$z\sim 0.5$ at R=22 to $z \sim 1$ at R=25 (Kneib et al. 1995).
There could however be a population even at redshift beyond 1,
as indicated by the detection of the shear around a 
$z=0.8$ cluster (Luppino \& Kaiser 1996). Most of the results we present
are for $z_s=1$, but they can be adapted to any desired $z_s$ using
the scalings of table 1. For simplicity we have assumed that all the
galaxies are at the same distance. This is the best possible case in the sense
that it makes the distributions in $k$ space more narrow than for more
realistic cases. Fortunately the difference between our 
results and those using a different, more 
realistic distribution of source galaxies with the 
same median redshift is usually very small (Kaiser 1992, Villumsen 1996). 

\begin{figure}[t]
\vspace*{9.3 cm}
\caption{The dependence of $\bar{p}$ on $\theta$
is shown for three cosmological models, with $z_s=1$ and a 
$\Gamma=0.25$ CDM spectrum. 
The thick solid curve is for $\Omega_m=1$, the dashed curve for 
$\Omega_m=0.3$ and the dotted curve for $\Omega_m=0.3$,
$\Omega_\Lambda=0.7$. The thin solid curve is for the linear 
spectrum with $\Omega_m=1$. The three panels show three alternative
normalizations of the power spectrum. In the case of COBE
normalization $\Omega_m=0.4$ has been used for the open and $\Lambda-$models;
see text for details of the other parameters for the COBE models. 
}
\includegraphics{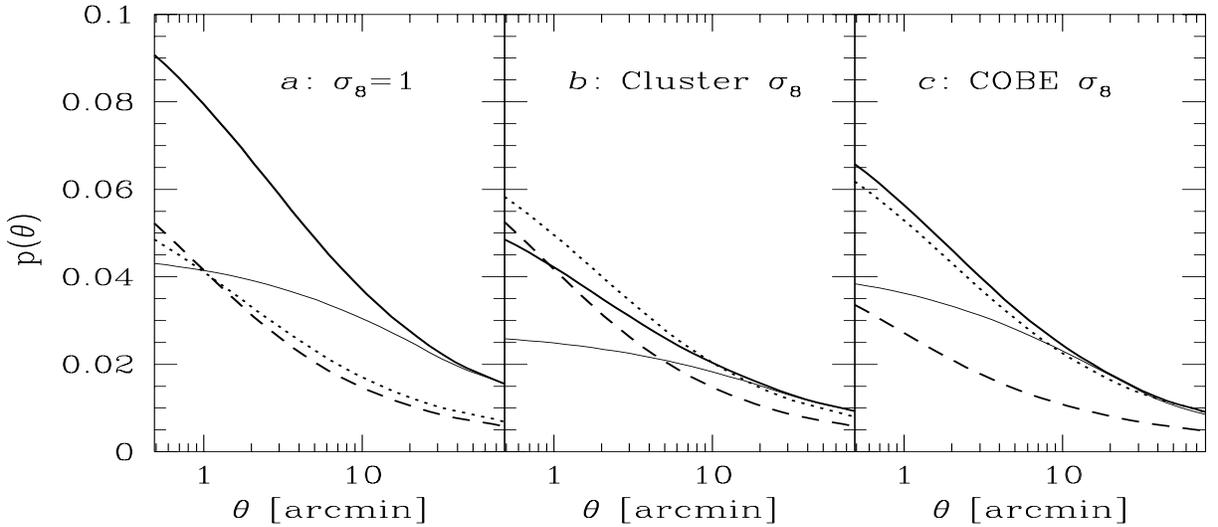}
\label{pvstheta1}
\end{figure}

Figure \ref{pvstheta1} shows the dependence of $\bar{p}(\theta)$ on $\theta$
for the three cosmological models, all with shape parameter 
$\Gamma=0.25$, except for the $\Omega_m=1$ COBE-normalized model. The thick 
solid curve is for $\Omega_m=1$, the dashed curve for 
$\Omega_m=0.3$ and the dotted curve for $\Omega_m=0.3$,
$\Omega_\Lambda=0.7$. The normalization chosen is (a)
$\sigma_8=1$, (b) Cluster-abundance, and (c) COBE. For the open
and flat COBE models, we have used $\Omega_m=0.4$ 
so that the value of $h$ and $\sigma_8$ are reasonable. 
The three COBE models are: flat, $\Omega_m=1$, tilted with $n=0.8$, 
$h=0.5$, $\sigma_8=0.72$; open with $\Omega_m=0.4$, $h=0.65$, 
$\sigma_8=0.64$; flat, $\Lambda-$model with $\Omega_m=0.4$, $h=0.65$, 
$\sigma_8=1.07$ (Bunn \& White 1996). Since the $\Lambda-$model 
chosen has a high $\sigma_8$, it turns out to have nearly the
same $\bar{p}(\theta)$ prediction as the tilted $\Omega_m=1$ model. 

The thin solid
curve is computed using the linear spectrum for the $\Omega_m=1$
model. As expected from the results of figure \ref{ratio}, it is only
for $\theta>10'$ that the linear and nonlinear curves start to
coincide. 
For cluster-abundance normalization, the amplitude of 
ellipticity correlations very weakly depends on the mean density for 
a reasonable range of $\Omega_m$.
The enhancement for low $\Omega_m$ 
models from the growth and distance factors discussed in \S 2
nearly cancels out the $\Omega_m^{0.4}$ dependence one might have
expected from the term outside the integral in equation \ref{p}. 
Therefore one has to make subtler comparisons of the signal at small and
large $\theta$ to break the degeneracy between $\sigma_8$ and
$\Omega_m$. This was discussed above 
following equation (\ref{scalings-lambda}). 
It indicates that comparing
weak lensing measurements over different scales can more powerfully 
constrain $\Omega_m$ than comparing the weak lensing amplitude with other 
tracers. For COBE normalization on the other hand, a detection of the 
correlated ellipticity
with reasonable error bars can immediately constrain the 
cosmological model in question, but of course there might still 
remain several different models which equally well fit the data. 

\begin{figure}[t]
\vspace*{9.3 cm}
\caption{The square root of the 
dimensionless ellipticity power spectrum $\sigma(1/\theta)$ 
is shown for the same three cosmological models as in figure
\ref{pvstheta1}. 
The left panel uses cluster-abundance normalized
spectra, while the right panel uses COBE normalized spectra. 
The thin solid curve shows $\sigma(1/\theta)$ for the $\Omega_m=1$ 
model computed using the linear power spectrum.
}
\includegraphics{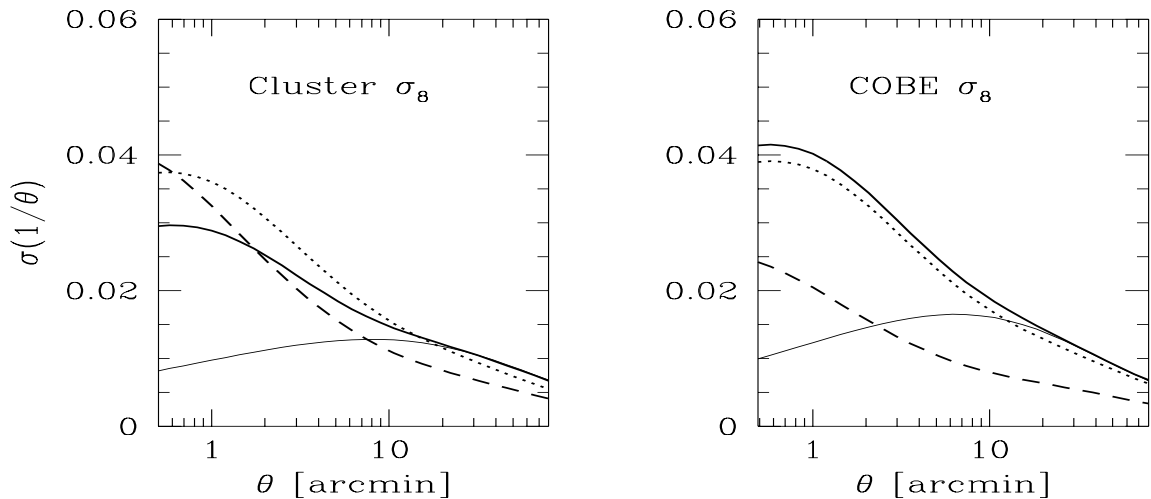}
\label{pvssig}
\end{figure}

\begin{figure}[t]
\vspace*{9.3 cm}
\caption{The effect of the shape of the power spectrum on 
$\bar{p}(\theta)\, $ is shown by comparing the $\Gamma=0.25$ CDM spectrum
(solid curves) with the $\Gamma=0.5$ CDM spectrum (dot-dashed curves). 
The upper set of curves is for $z_s=2$, and the lower set for
$z_s=1$. The $\Gamma=0.5$ spectrum has more small scale power and 
therefore predicts larger $\bar{p}$ at small $\theta$. 
}
\includegraphics{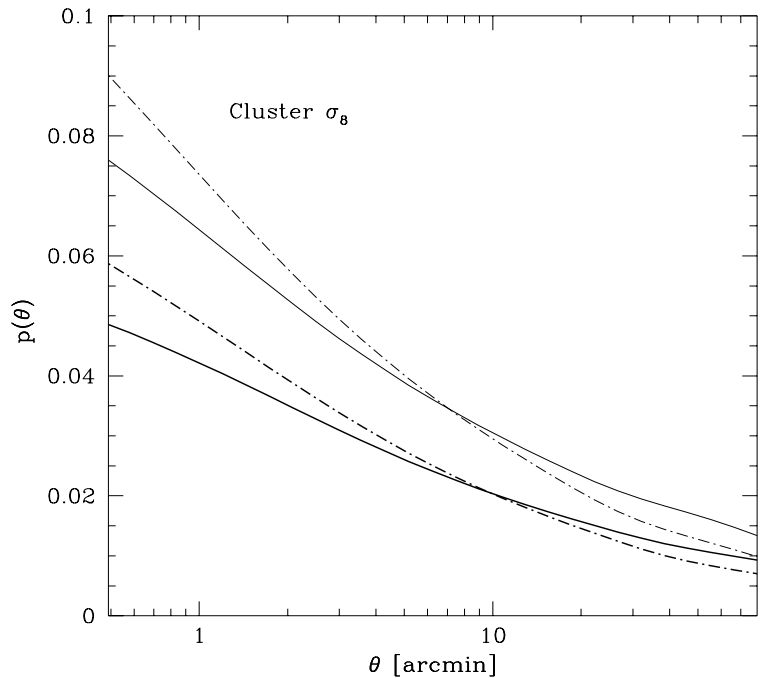}
\label{shape}
\end{figure}

\begin{figure}[t]
\vspace*{9.3 cm}
\caption{The effect of increasing source redshift $z_s$ on
$\bar{p}(\theta)\, $ is shown for $\theta=2'$ (upper 3 curves)
and $\theta=15'$ (lower curves). 
The three models shown are the same as in figure \ref{pvstheta1}
and use cluster-abundance normalized spectra. The $\Lambda-$ models, 
shown by the dotted curves, predict the fastest growth with $z_s$ as discussed
in the text. 
}
\includegraphics{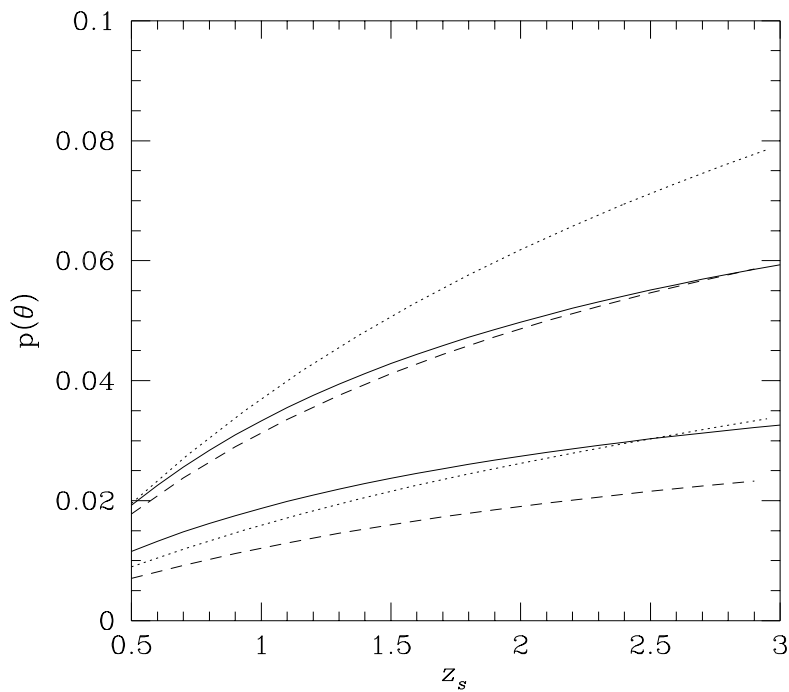}
\label{zs}
\end{figure}

The effect of choice of statistic, shape of the power spectrum, and
the redshift of source galaxies is shown in figures \ref{pvssig}
and \ref{shape}. Figure \ref{pvssig} shows $\sigma(1/\theta)$
for the three cosmological models with 
cluster-abundance and COBE normalization as in figure \ref{pvstheta1}. 
The curves for $\sigma$ flatten at small $\theta$ more than 
the $\bar{p}$ curves. 
This is to be expected because $\bar{p}$ integrates over all scales
larger than $r\theta$ (where $r$ is the typical distance) and so 
can only increase toward small scales. Because there is little power 
on very small scales it eventually saturates (this typically happens
on subarcminute scales, Keeton, Kochanek \& Seljak 1996).
The Fourier space quantity $\sigma^2(l)$ on the other hand receives
contributions only from scales smaller than $r/l$
and therefore decreases on very small scales for realistic power 
spectra.
The shape of the $C_{pp}$ curve is very similar to that of $\bar{p}^2$, 
and for $\theta\gsim 2'$, $C_{pp}$ is well approximated by $0.7\, \bar{p}^2$. 
Figure \ref{shape} compares $\bar{p}(\theta)$ 
for the $\Gamma=0.5$ CDM spectrum
to the $\Gamma=0.25$ spectrum used in the rest of the figures. The
upper and lower set of curves are for $z_s=2$ and $1$ respectively. 
Since the $\Gamma=0.5$ spectrum has more power at small scales, 
it leads to a larger $\bar{p}$ at small $\theta$. The
curves cross around $\theta=10'$, but the difference is not large 
for $\theta>2'$. Figure \ref{zs} shows the variation of $\bar p(\theta)$
with $z_s$ for $\theta=15'$ and $\theta=2'$. The models shown are
the same as in the cluster-abundance normalized panel of Figure
\ref{pvstheta1}. As shown in Table 1, $p$ increases with 
$z_s$ fastest for the $\Lambda-$models. 

\section{Nonlinear effects on the distribution of ellipticity}

So far we have only discussed the importance of nonlinear effects on the 
second order statistic, which as we have seen increases the 
signal above the values predicted by linear calculations. Nonlinear 
evolution however also changes the distribution function of correlated
ellipticity or 
magnification. 
In particular, the tails of distribution grow and become asymmetric, 
reflecting the nonlinear growth of perturbations  
into small and overdense objects. 
This leads to a distribution of magnification
which is skewed toward positive values. Such skewness is of 
interest for determining cosmological parameters by itself 
(Bernardeau et al. 1996), but also affects the
inference of cosmological parameters from the 
measurements of 2-point statistics.
If the distribution of polarization at a given angular scale 
$\theta$ can be assumed to be normal distributed
then the usual results from 
gaussian statistics apply. 
For example a single measurement of polarization amplitude within a 
circular aperture has a Rayleigh probability distribution 
$\chi^2(2)$, which has exponentially suppressed tail and from 
which the limits on cosmological parameters can be derived analytically. 
Several independent measurements can be 
easily combined and only a few observations suffice to obtain strong
limits on cosmological parameters. 
If, on the other hand, 
the distribution is strongly non-gaussian so that the tails of 
distribution are important
then one cannot use these simple arguments to derive the limits and 
a larger set of measurements (a "fair sample")
is needed to sufficiently sample the tails
so that the variance converges to a true value. 
At the same time, higher order statistics become easier to measure
and can provide further tests of cosmological parameters.

For measurements of density perturbations a simple criterion of where
the nonlinear effects become important is $\sigma_R\sim 1$, so for scales 
above $R \sim 10h^{-1}$ Mpc the universe is in the linear regime and gaussian
statistic can be used. For smaller scales nonlinear effects become 
important which leads to deviations from a normal distribution even if the
initial field was gaussian. 
For projected quantities
such as shear and magnification the answer is not so simple. The
projection integrates over all the scales, so there will always be 
contribution from very small and thus very nonlinear scales.  
On the other hand, projection will reduce the deviations from a gaussian 
distribution because of the central limit theorem: if we sum over 
sufficiently large number of deflectors 
the final distribution will be a gaussian regardless
of the probability distribution of individual deflectors. We thus expect
that the nonlinear effects will be more important for small angles and 
for bright (nearby) background galaxies. 

\begin{figure}[p]
\vspace*{10.3 cm}
\caption{Skewness $S=\langle \bar{\kappa}^3\rangle/\langle \bar{\kappa}^2 
\rangle^{3/2}$ as a function of $\theta$ for the flat model 
(solid curves), open model 
(dashed curves) and cosmological constant model (dotted curves). 
The upper set of curves is for $z_s=0.5$, middle for $z_s=1$ and 
bottom for $z_s=2$. Over most of the regime of interest the nonlinear 
effects are important. Note that for $\theta\lsim 15'$ perturbation
theory underestimates the skewness of the density (Colombi et
al. 1996),  so that on arcminutes scales $S$ could be nearly a factor
of 2 larger than shown in this figure. 
}
\label{skew1}
\includegraphics{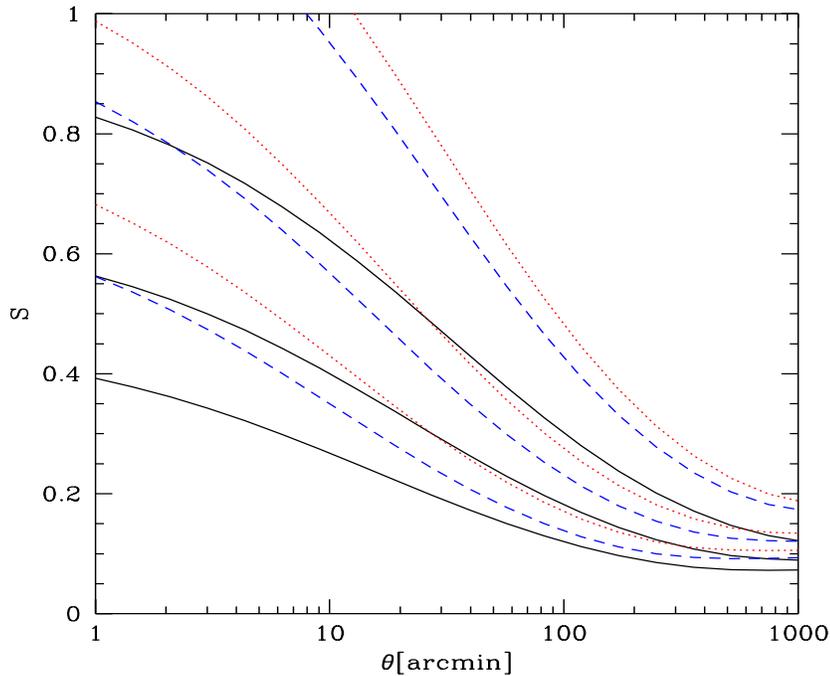}
\end{figure}

To obtain a more quantitative answer we calculated the skewness of 
local convergence 
$S=\langle \bar{\kappa}^3\rangle/\langle \bar{\kappa}^2 \rangle^{3/2}$ 
using second order perturbation theory. 
Skewness for local convergence does not vanish as 
in the case of ellipticity (Bernardeau et al. 1996).
The second moments of 
$2\bar{\kappa}$ and $\bar{p}$ are the same
and so one may expect the higher order moments of convergence to 
be indicative of the nonlinear effects on the ellipticity distribution as well.
Note that the quantity $S$ differs from the quantity 
$s_3=\langle \bar{\kappa}^3\rangle/\langle
\bar{\kappa}^2 \rangle^2$ used by Bernardeau et al. 1996. The latter 
is defined so that it is independent of the amplitude of power spectrum 
in the second order perturbation theory, while $S$ is a more useful variable
to quantify the deviations from the gaussianity. 
As discussed in the appendix, 
second order perturbation theory should be approximately valid 
in the regime of interest here, between 1' to 1$^\circ$. 
Second order perturbation theory accurately estimates the skewness 
over a much wider range of scales than it does the variance. 
Only on the smallest scales it underestimates the skewness, but typically 
not more than a factor of two, as shown by
N-body simulations (Colombi, Bouchet \& Hernquist 1996). 
Using the expressions derived in appendix (see also Bernardeau et al. 1996)
we can compute the value of $S$ for any of the models 
discussed in this paper. 
Figure \ref{skew1} presents the distribution of $S$ as a function of
$\theta$ for 3 different redshifts in 3 different models: flat model 
with $\sigma_8=0.6$, curvature and cosmological constant models, 
both with $\sigma_8=1$ and $\Omega=0.3$ (this is very close to the 
cluster-abundance normalization). In all the models we use
$\Gamma=0.25$.

For simplicity we will choose $S=0.2$ as the value where the distribution 
becomes non-gaussian. For background galaxies at $z_s=0.5$ we find that 
the distribution is non-gaussian even at $4^\circ$ in the flat model, 
where the 
polarization signal is only 0.25\%. At $z_s=1$ and $z_s=2$ the corresponding 
values are $1^\circ$ and $0.5^\circ$.  
For curvature and cosmological constant 
dominated models the non-gaussian effects are even 
more important because of higher normalization at $8 h^{-1}$Mpc, although the 
effect is offset by the longer radial distance to a given redshift. 
The conclusion is that in most of the observationally interesting 
regime 
the nongaussian effects will be quite important. While this 
complicates the issue of how large a fair sample should be, it 
also implies that skewness will be fairly easy to measure with 
a reasonably small sample. Because $s_3$ 
depends sensitively on $\Omega_m$ 
but not on the amplitude of the power spectrum (Bernardeau et al. 1996)
this test may become a promising way to determine $\Omega_m$. 

The analytical approach  presented above 
provides only a limited view of the properties of ellipticity distribution.
For example, such an analysis cannot answer whether 
skewness is produced by a few outlying points
generated by rare clusters, in which case a rather large
sample would be needed to sample these properly.
For a quantitative estimate of what
constitutes a fair sample for measuring the variance requires therefore
knowledge of at least the fourth moment. Such estimates of a fair
sample for measuring the second and third moment can be made more 
directly by examining the distribution of correlated ellipticities
in N-body simulations (Wambsganss, Cen \& Ostriker 1996). 

\section{Discussion and conclusions}

Weak gravitational lensing has several advantages over other probes of
large scale structure. It is directly sensitive to the underlying 
mass distribution and is therefore independent of biasing, which 
has plagued many of the cosmological tests based on the distribution
of galaxies. While peculiar velocities in principle similarly probe
the dark matter distribution directly, in practice they suffer because
of large observational errors and various biases associated with these. 
This is the main reason why higher order statistics of velocity 
fields have not provided
strong constraints on cosmological models. 

\begin{figure}[t]
\vspace*{9.3 cm}
\caption{The quantity $(\theta/5')\, \bar{p}^2(\theta)\, 10^5$ 
is shown for the three models of figure \ref{pvstheta1}. This provides
an estimate of the signal to noise in the measurement of $\bar{p}^2$ in 
a one square degree field with $2\times 10^5$ galaxies, as obtained from a 
typical observation with a limiting magnitude $I=26$. 
Nonlinear evolution causes the curves to be much less sensitive
to $\theta$ in the range $2'<\theta<2^\circ$ compared to the 
linear prediction, which shows a significant peak around $\theta=1^\circ$.
The linear curves for the three models are shown by the thin curves. 
}
\includegraphics{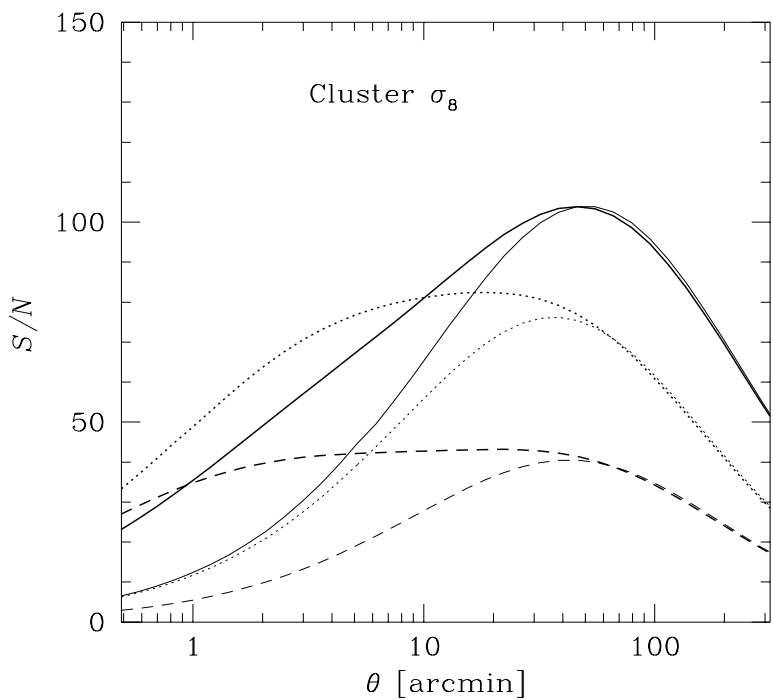}
\label{sn}
\end{figure}

Gravitational lensing could
provide a much cleaner way of performing power spectrum and higher order
statistical tests, provided that one can 
overcome observational difficulties and reach the 1\% level of  
polarization, quite feasible with the new generation of composite
CCD cameras. For a filled survey with an area $\theta_0^2$ and 
$N$ measured galaxies the noise variance in $\bar{p}^2(\theta)$ is given by 
$\sim (0.4)^2\theta_0/N\theta$, where $0.4$ is the typical intrinsic 
shear for a single galaxy. 
Although one has to average over sufficient number of
galaxies to reach this level of signal,
this is not a severe limitation for sufficiently deep exposures with
several hundred thousand galaxies per square degree, so on degree
scales the signal to noise 
is of the order of 100 (Kaiser 1996; fig. \ref{sn}) and in fact 
sampling variance (finite number of independent areas of 
size $\theta^2$) is more important than noise on large angles. On 
smaller scales the noise increases (proportional to $\theta^{-1}$), 
but the signal also increases nearly as rapidly, so that
the overall signal to noise remains approximately
flat down to $2'$ scales in open and cosmological constant models and
slowly decreases in flat models (fig. \ref{sn}). Based on this 
it appears there 
should be sufficient power present in a survey of this size
to measure the signal even on 
arcminute scales with a high statistical accuracy, so a comparison 
between small and large scales to constrain $\Omega_m$
would indeed be feasible, 
provided that the systematic effects can be kept under control. 

The other limitation at present is that the redshift distribution
of background galaxies is unknown beyond $I=23$ (Lilly et al. 1996; 
Cowie et al. 1996). 
New observations with Keck will
be able to extend this spectroscopically up to $I=25$ 
(e.g. Lowenthal et al. 1996) and even beyond,
if spectra of magnified arcs 
are measured. In addition,
weak lensing on high redshift clusters can be used to 
constrain the redshift 
distribution of faint galaxies 
without measuring their redshifts, although at present there appears
to be some controversy over whether the faintest galaxies are predominantly 
around or below $z=1$ (Kneib et al. 1995) 
or above that (Luppino \& Kaiser 1996). 
Finally, the best hope to constrain their redshift distribution may lie
in the photometric redshifts technique 
(Conolly et al. 1995; Sawicki, Lin \& Yee 1996).
This would allow one to perform such precision tests as 
the growth of power spectrum with redshift, needed to estimate 
the value of $\Omega_\Lambda$. For example, based on Hubble Deep Field 
observations, Sawicki et al. 1996 find that the redshift distribution of 
$I<27$ galaxies is bimodal, with a large contribution of (predominantly
blue) galaxies between redshifts 2-3, while at $I<25.5$ most of the 
galaxies are below $z=1$. In both cases the weak lensing 
signal to noise with these limiting magnitudes would be comparable to 
the one in fig. \ref{sn},
so in principle by comparing the signals in the two magnitude
bins it would be straightforward to deduce the value of 
$\Omega_\Lambda$. The caveat is that the faintest galaxies in HDF appear
to be very irregular and small, so it is not clear whether weak lensing 
analysis from the ground could be performed on these objects. While 
more investigation is needed along these lines, there appears to be 
a real promise of measuring cosmological parameters with 
future weak lensing observations.

We have analyzed theoretical predictions for weak lensing 
ellipticity correlations over a broad range of angles and 
for a variety of cosmological models. Our principal conclusions
are the following: 

a) the amplitude of the rms ellipticity 
varies between 4-6\% on arcminute scales to 0.5-1\% on degree
scales for background galaxies at $z_s=1$ and grows
as $z_s^{0.6-0.8}$. In this range the amplitude only weakly depends on 
the value of cosmological parameters if cluster abundance
(or velocity flows) 
normalization is adopted. A single measurement of weak 
lensing amplitude on $\theta\simeq 15'$ will provide a 
determination of $\sigma_8\, \Omega_m^{0.7-0.8}$ which can be 
compared with the cluster-abundance or velocity flow predictions. 
Because these tracers all scale similarly with $\Omega_m$ this will
not provide an independent determination of
$\Omega_m$, for which a more complicated analysis will be needed.

b) nonlinear effects are important for second order statistics
on scales below $10'$ and enhance the signal expected from linear 
theory. This enhancement can be up to a factor of two on arcminute scales. 
 From this it follows that the signal to noise does not decrease
as rapidly on small scales as expected from linear theory
and is in fact approximately flat in certain models 
from arcminute to degree scales (see figure \ref{sn}). 
Therefore measurements over this whole angular range
should be pursued, especially since they provide complementary 
information. The interpretation of cosmological models
is more cleanly addressed on large angular scales, where nonlinear 
effects are negligible and the distribution function is 
primordial. Small angular scales can provide insight on the 
nonlinear evolution of perturbations. If our current understanding of the
latter is correct then combining measurements on large and small
angular scales will help break the degeneracy between 
the cosmological parameters, because of different scalings in 
the linear and nonlinear regimes. 
The ellipticity on small angles measures $\sigma_8 \Omega_m^{0.5}$, on
large angles it measures $\sigma_8 \Omega_m^{0.8}$, while the 
variation with $z_s$ can constrain $\Omega_\Lambda$. 
Thus, to obtain model independent measurements of 
the cosmological parameters  it is necessary to
measure the signal over a range of angles and source redshifts. 

c) nonlinear effects on the distribution of ellipticity 
are significant on scales below $1^\circ$ and imply that
gaussian statistics cannot be applied to the data. This means that 
a larger sample will be required to measure second order 
statistics than expected from a gaussian distribution.   
Conversely, skewness and other higher moments will be 
easy to measure and may provide a strong test of 
cosmological parameters independent of the tests above.

In conclusion, weak lensing has the promise to become the 
next testing ground for cosmological theories. The expected signal to noise
is above unity over a large range of angles and limiting magnitudes. This 
should allow one to test sensitively the spectrum of density perturbations
and its evolution in time, in both the linear and the nonlinear regimes. 

\acknowledgements
B.J. is very grateful to Jens Villumsen for initiating his interest
in weak lensing, and to him and Richhild Moessner for useful
discussions. We thank Bernard Fort, Nick Kaiser, Chris Kochanek,
Yannick Mellier, Stella Seitz, 
Simon White, and especially Peter Schneider for stimulating
discussions. We thank the referee for suggestions which helped
improve the paper and led to the addition of figure 8. 

\appendix
\section{Appendix}

In this Appendix we present a general description of lensing 
in a weakly perturbed universe. The derivation presented here follows
Seljak (1995, 1996) and differs from recent derivations in Bernardeau et al.
(1996) and Kaiser (1996) in that it employs a global description of 
photon trajectories. 
The geometrical interpretation requires only knowledge of
spherical trigonometry and it is thus particularly simple to derive the
general weak lensing expressions.
Moreover, it can easily be generalized beyond  
weak lensing approximation or small deflection angles. 

The framework is a perturbed Robertson-Walker
with small-amplitude metric fluctuations. We will only consider 
scalar perturbations here, in which case the metric in the longitudinal 
gauge can be written using conformal time $\tau$ and comoving
coordinates $x^i$ as 
\begin{equation}
ds^2=a^2(\tau)\left\{-(1+2\phi)d\tau^2+(1-2\phi)
    \gamma_{ij}dx^idx^j\right\}.
    \label{metric}
\end{equation}
Here $a(\tau)$ is the scale factor expressed in terms of conformal
time. We will adopt units such that $c=1$. We set the two scalar 
potentials to be equal, which is a good approximation in the matter 
dominated epoch. The space part of the background metric can be written as
\begin{eqnarray}
\gamma_{ij}dx^idx^j=d\chi^2+r^2(d\theta^2+\sin^2 \theta d\phi^2),
\nonumber \\
\nonumber \\
r(\chi)=\sin_K\chi \equiv
\left\{ \begin{array}{ll} K^{-1/2}\sin K^{1/2}\chi,\ K>0\\
\chi, \ K=0\\
(-K)^{-1/2}\sinh (-K)^{1/2}\chi,\ K<0\\
\end{array}
\right.
\label{rchi}
\end{eqnarray}
where $K$ is the curvature term which
can be expressed using the present density
parameter $\Omega$ (only the present day density parameter 
will be used in this paper, hence we may drop the subscript 0)
and the present
Hubble parameter $H_0$ as $K=(\Omega-1)H_0^2=-\Omega_KH_0^2$.
The relation between the radial distance $\chi$ and the 
redshift $z$ can be obtained
from the Friedmann equation, $da/d\chi=H_0(\Omega_{m}a+
\Omega_{\Lambda } a^4 + \Omega_Ka^2)^{1/2}$, 
where $a=(1+z)^{-1}$ is the expansion factor and the densities
of matter, cosmological constant and curvature
are expressed in terms of the critical density. 
The density parameter $\Omega$ can have contributions from mass
density
$\Omega_{m}$ or
vacuum energy density $\Omega_{\Lambda}$, $\Omega=\Omega_{m}+
\Omega_{\Lambda}$.
The advantage of using the conformal time $\tau$ is that the metric
becomes conformally Euclidean ($K=0$), 3-sphere ($K>0$) or
3-hyperboloid ($K<0$) and leads to a simple geometrical description
of light propagation.

\begin{figure}[p]
\vspace*{6.0 cm}
\caption{Photon propagation relative to the source-observer line: a
photon is emitted at the source and observed at the observer's
position in the direction $\vec \theta$ relative to the unperturbed
source-observer direction (which is curved on an Euclidean plane because of
background curvature).
A deflection at $\chi'$ by $\delta \vec \alpha=-2\vec \nabla_\perp \phi
\delta \chi$ leads to the transverse excursion at $\chi$ given by
$\vec \delta x_\perp(\chi)=
r(\chi - \chi')\delta \vec \alpha$.
}
\label{fig_appb}
\includegraphics{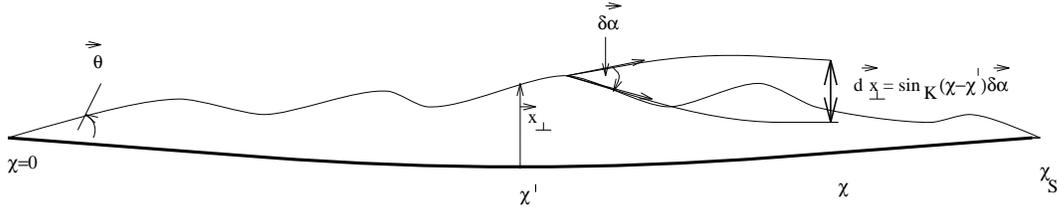}
\end{figure}

In an unperturbed universe a photon emitted from a source
toward the observer will travel along a
null geodesic in the radial direction with a radial position given
by $\chi=\tau_0-\tau$. Adding a perturbation changes the photon trajectory.
The change in photon
direction is governed by the space part of the geodesic equation, which 
applied to the metric (equation \ref{metric}) gives
\begin{equation}
{d\vec n \over dl}=2\vec n \times (\vec n \times \vec \nabla
\phi)\equiv -2 \vec \nabla_\perp \phi,
\label{dndl}
\end{equation}
where the symbol $\vec \nabla_\perp \phi$ denotes the transverse
derivative of potential.
Null geodesics obey $ds^2=0$, from which follows
the relation $d\chi=(1-2\phi)dl$.
Here $\phi$ can be interpreted as the Newtonian potential, since on
scales smaller than the horizon it
obeys the cosmological Poisson equation. 
It can be viewed as providing a force deflecting the photons
and affecting their
travel time while they
propagate through the
unperturbed space-time. 
Equation (\ref{dndl}) is a generalization of Einstein's deflection
angle formula and includes the well-known factor of 2 difference
compared to Newtonian gravity. Even when
metric perturbations are present, one can
continue to parameterize the geodesic with the unperturbed comoving
radial distance $\chi$.
The deflection angle at a given position $\chi$ can
be calculated using a locally flat coordinate system,
which allows a plane wave expansion for the potential $\phi$,
provided that the longest correlation length is small compared to the
curvature length (this condition is 
well satisfied for the power
spectrum in our universe).
The effect of the deflection angle on the
photon transverse position must however include the curvature effects. In
practice this means that one only needs to know how to solve triangles
using the spherical,
Euclidean or hyperboloid trigonometry (figure \ref{fig_appb}).
Because the only observable photon direction is that at the observer's
position we will propagate photons relative to their final direction
(i.e. backwards in time).
We will also adopt a small deflection angle approximation, because one
does not expect large deflection angles due to the lensing, but the 
expressions can be generalized to remove this restriction.
In this approximation the transverse derivatives in equation (\ref{dndl}) can be approximated
with the transverse derivatives with respect to the observed direction
of the photon or with respect
to any other fiducial direction that has a small angular
separation with the photon (for example the unperturbed
direction to the source).
In this plane approximation
the observed photon direction $\vec n$ can be described with a two-dimensional
angle $\vec \theta$ with respect to the fiducial direction, $\vec n=(\theta_
1,
\theta_2,1-\vert \vec \theta \vert ^2/2\approx 1)$.

Suppose a photon is observed at an angle $\vec \theta$ relative to some
fiducial position.
As it propagates through the universe the photon is additionally
deflected according to equation
(\ref{dndl}) (see figure \ref{fig_appb}).
This leads to the transverse photon excursion $\vec x_\perp(\chi)$
relative to the unperturbed line.
 From Euclidean, spherical or hyperboloid
trigonometry one finds
that an individual deflection by $\delta \vec \alpha$
at $\chi'$ leads to an excursion at $\chi$ given by
$\delta \vec x_\perp=\delta \vec \alpha 
r(\chi-\chi')$. The total excursion is given by an integral over individual
deflections,
\begin{equation}
\vec x_{\perp}(\chi)=-2\int_0^\chi \vec\nabla_\perp\phi(\chi')r(\chi-
\chi')d\chi'+\vec \theta r(\chi).
\label{xperp}
\end{equation}

Because of gravitational lensing the
``true'' surface brightness (i.e. the surface brightness one would see in the
absence of any lensing) at position $\chi_S$ is mapped into the observed one,
$I_{{\bf obs}}(\vec \theta)=I_{{\bf true}}(\vec \theta +\delta \vec \theta)$,
where $\delta \vec \theta$ is
the angular deflection of a photon caused by intervening mass, which follows
from equation (\ref{xperp})
\begin{equation}
\delta \vec \theta={-2 \over r({\chi_S}) } \int_0^{\chi_S}r(\chi_S
-\chi)
\vec \nabla_\perp \phi(\chi) d\chi.
\label{dtheta}
\end{equation}
While the deflection $\delta \vec \theta$
is not directly observable, its gradient is,
through the stretching and magnification of distant galaxies.
This is described by the two-dimensional shear tensor defined in 
equation (\ref{shear}).
To compute it we expand the potential in equation (\ref{dtheta})
across two neighboring rays separated at the observer
by $d\theta$ using the {\it unperturbed} separation
$dx=r(\chi) d\theta$.
This approximation assumes that the components of shear tensor 
$\Phi_{ij}$ are small (Kaiser 1992),
or, similarly, that the relative deflection between the neighboring
rays is small compared to the unperturbed separation. This is the so-called
weak lensing approximation. 

The shear tensor can be diagonalized and decomposed 
into its trace $2\kappa=\Phi_{11}+\Phi_{22}$ 
and ellipticity (or polarization) $\bi{p}=p_1+ip_2$ given by 
\begin{equation}
p_1=\Phi_{11}-\Phi_{22}\ ; \ p_2=2\Phi_{12}\, .
\label{shear3}\end{equation}
While trace $2\kappa$ magnifies (or demagnifies)
the images of  
galaxies, ellipticity $\bi{p}$ stretches
the images and can be observed by averaging over a sufficient number of
galaxies so that the noise caused by the intrinsic ellipticity of galaxies
becomes smaller than the signal. The simplest average is the one
within a circular window of radius $\theta$, 
which can be written in Fourier space as 
\begin{equation}
\bar{\bi{p}}(\theta)={2 \over \pi \theta^2}
\int_0^\theta d^2\vec{\theta}' \int_0^{\chi_0}g(\chi)d\chi
\int d^3\vec{k} e^{ik_\chi\chi}e^{ir\vec{k}_\perp \cdot \vec{\theta'}} 
k^2_{\perp } \bi{\beta}\phi(\vec{k}).
\label{a1}
\end{equation}
We decomposed the wavevector into the radial component $k_\chi$ and 
the 2-dimensional transverse component $\vec{k}_\perp$ with 
the amplitude $k_{\perp}$. We introduced the variables $\bi{\beta}=
exp(2i\phi_k)$, where $\phi_k$ is the azimuthal angle of 
$\vec{k}_\perp$. The
integral over $d^2\vec{\theta}'$ can be readily performed to give
\begin{equation}
\bar{\bi{p}}(\theta)=2\int_0^{\chi_0}g(\chi)d\chi \int d^3\vec{k} 
e^{ik_\chi \chi}k^2_{\perp } \bi{\beta} \phi(\vec{k})
W_2(k_{\perp}r\theta),
\end{equation}
where $W_2(x)=2J_1(x)/x$.
We may now employ the small angle approximation (Limber 1956), 
which is valid in the 
limit where the radial window function is broad compared to the typical 
wavelength that contributes to the integral. In this limit only the modes
perpendicular to the radial direction will contribute to the integral, 
all the others being suppressed because of cancellation of positive
and negative fluctuations along the line of sight. This allows one to 
put $k_\perp \approx k$. With this approximation one obtains 
for rms ellipticity amplitude $\bar{p}^2=\bar{\bi{p}}\bar{\bi{p}}^*$
\begin{equation}
\langle \bar{p}^2(\theta) \rangle = 4 \int \int d^3\vec{k}_1 d^3\vec{k}_2 
\int_0^{\chi_0}
\int_0^{\chi_0}d\chi_1 d\chi_2 e^{i(k_{\chi_1}\chi_1+k_{\chi_2}\chi_2)}
W_2(k_1r_1\theta)W_2(k_2r_2\theta)k_1^2k_2^2 \langle \phi(\vec{k_1})\phi(\vec{k_2})\rangle.
\end{equation}
Ensemble average can be expressed in terms of power spectrum of potential 
$P_\phi(k)$ or density $P_\delta(k)$,
\begin{equation}
\langle \phi(\vec{k_1})\phi(\vec{k_2})\rangle= P_\phi(k_1,a)\delta_D(\vec{k_1}+\vec{k_2})
={9 \over 4}\left({\Omega_m \over a}\right)^2 \left({H_0\over k}\right)^4P_\delta(k_1,a)\delta_D(\vec{k_1}+\vec{k_2}),
\end{equation}
where power spectrum $P_\delta(k,a)$ and expansion 
factor $a$ depend on time. 
Integrating over the 
Dirac $\delta_D$ function and over the radial component of wavevector
we finally obtain equation (\ref{p}), which gives the rms amplitude 
of polarization. The same expression can also be used for
rms amplitude of magnification $2\bar{\kappa}$. 
The rms amplitude for each of 
the two ellipticity components is simply
$1/\sqrt{2}$ of the total. 

Because skewness of ellipticity vanishes for symmetry reasons
we will use skewness of local convergence as an estimate of the nongaussian 
nature of ellipticity distribution.
Calculating it   
proceeds along the same lines as for the variance (Bernardeau 1995). 
It is defined as an ensemble average of the third moment of the mean 
magnification,
\begin{equation}
\langle \bar{\kappa}^3(\theta)\rangle=\left\langle \left[ 
\int_0^{\chi_0}d\chi\int d^3\vec{k}  e^{ik_\chi \chi}
g(\chi) W_2(k_\perp r \theta) k^2 \phi(\vec{k}) \right]^3
\right\rangle.
\end{equation}
In linear theory the 3-point correlation function vanishes, so one needs
to go beyond the linear approximation to obtain a nonvanishing value. 
In second order perturbation theory the density field is given by
\begin{eqnarray}
\delta^{(2)}(\vec{k})&=&\int d^3\vec{k}_1 d^3\vec{k}_2 
\delta^{(1)}(k_1)\delta^{(1)}(k_2)
\delta_D(\vec{k}_1+\vec{k}_2-\vec{k})F(\vec{k}_1,\vec{k}_2) \nonumber \\
F(\vec{k}_1,\vec{k}_2)&=&\left[{5 \over 7} +
 {\vec{k_1} \cdot \vec{k_2} \over  k_1^2}
+{2 \over 7} {(\vec{k_1} \cdot \vec{k_2})^2 \over k_1^2 k_2^2 }
\right].
\end{eqnarray}
Note that the time dependence is all in the growth factors of 
$\delta(k)$, which is calculated using linear theory
(neglecting the extremely weak dependence of $F(\vec{k}_1,\vec{k}_2)$ 
on $\Omega$, Bouchet et al. 1992).
Ensemble averaging of skewness gives 6 identical terms, which after 
integrating over the radial wavevectors leads to the following 
expression,
\begin{eqnarray}
\langle \bar{\kappa}^3(\theta)\rangle&=&81\pi^2\theta^{-4}H_0^6
\int_0^{\chi_0} g^3D^4\Omega^3a^{-3}r^{-4}d\chi \nonumber \\
&\times& \int d^2\vec{l}_1 d^2\vec{l}_2
W_2(l_1)W_2(l_2)W_2(|\vec{l}_1+\vec{l}_2|)
P_\delta(l_1/r\theta)
P_\delta(l_2/r\theta)F(\vec{l}_1,\vec{l}_2).
\end{eqnarray}
The
integrals can be further simplified using similar manipulations
as in Bernardeau (1995), which leads to the expression 
(see also Bernardeau et al. 1996)
\begin{eqnarray}
\langle \bar{\kappa}^3(\theta)\rangle&=&
322\pi^4\theta^{-4}H_0^6
\int_0^{\chi_0} g^3D^4\left({\Omega\over a}\right)^3r^{-4}d\chi 
\int P(l/r\theta)W_2^2(l)ldl
\nonumber \\ &\times& 
\left[{6 \over 7}\int P(l/r\theta)W_2^2(l)ldl
+{1 \over 2}\int P(l/r\theta)W_2(l)W_2'(l)l^2dl \right].
\label{skew}
\end{eqnarray}
This expression was used in \S 4 to evaluate the importance of 
nonlinear evolution on the distribution function of polarization.

Several approximations have been employed in obtaining the result above,
which limit its validity. 
Second order perturbation theory 
results are only valid in the domain where density perturbations are 
not much larger than unity, 
so on very small scales
a more involved calculation would be needed. However, these
corrections are typically not more than a factor of two. Because our 
purpose is to estimate at which scale the nongaussian effects become
important second order perturbation theory suffices.
Another approximation is the use of small angle
approximation, which leads to corrections on  
angular scales above $1^\circ$ 
(Bernardeau 1995). In the regime of main interest for us, which is between 
$1'-1^\circ$ this is not an important correction.
Finally, expression (\ref{skew}) neglects 
the deviation of the photon trajectory from the unperturbed path, which
also contributes at the second order. This correction is  
less important than the main term used in equation (\ref{skew})
and can be neglected (Bernardeau et al. 1996).

\end{document}